\def   \ni {\noindent}

\def   \sssk {\vskip  0truept} 
\def   \ssk {\vskip  5truept}
\def   \sk  {\vskip 10truept}
\def   \bsk {\vskip 15truept}
 
\def   \newpage {\vfill\eject}
\def   \newline {\hfil\break}
\def   \ergs {$erg~cm^{-2}~s^{-1}~$}
\def   \sax {{\it Beppo}SAX$~$}

\documentstyle[epsfig]{article}
\begin{document}

\hsize 5truein
\vsize 8truein
\font\abstract=cmr8
\font\keywords=cmr8
\font\caption=cmr8
\font\references=cmr8
\font\text=cmr10
\font\affiliation=cmssi10
\font\author=cmss10
\font\mc=cmss8
\font\title=cmssbx10 scaled\magstep2
\font\alcit=cmti7 scaled\magstephalf
\font\alcin=cmr6 
\font\ita=cmti8
\font\mma=cmr8
\def\ref{\par\noindent\hangindent 15pt}
\null


\title{\ni X-RAY SURVEYS}
\bsk \bsk
\author{\ni P. Giommi $^{1}$, F. Fiore $^{1,2,3}$ and M. Perri $^{1}$ }     
\ssk
\affiliation{1) \sax Science Data Center, Rome, Italy \sssk
2) Osservatorio Astronomico di Roma, Monteporzio, Italy \sssk
3) Smithsonian Astrophysical Observatory, 60 Garden Street, Cambridge, MA 02138
}                                                
\bsk
\baselineskip = 12pt

\abstract {ABSTRACT \ni
A review of recent developments in the field of X-ray surveys,
especially in the hard (2-10 and 5-10 keV) bands, is given. 
A new detailed comparison between the measurements in the hard band and 
extrapolations from ROSAT counts, that takes into proper account the observed 
distribution of spectral slopes, is presented. 
Direct comparisons between deep ROSAT and \sax images show 
that most hard X-ray sources are also detected at soft X-ray energies. 
This may indicate that heavily cutoff sources, that 
should not be detectable in the ROSAT band but are expected in large numbers 
from unified AGN schemes, are in fact detected because of the emerging of 
either non-nuclear components, or of reflected, or partially transmitted 
nuclear X-rays. These soft components may complicate the estimation 
of the soft X-ray luminosity function and cosmological evolution of AGN.
}                                                    
\ssk
\baselineskip = 12pt
\baselineskip = 12pt


\text{\ni 1. INTRODUCTION
\ssk
Over the past 30 years many X-ray surveys have been 
successfully carried out with the aim of both discovering new types 
of X-ray emitters and to investigate the nature of the historically first,
and still very popular, source of extragalactic X-rays: the cosmic X-Ray 
Background (XRB). 
It is now widely accepted that most, and probably all, the XRB can be 
explained as the superposition of faint discrete sources, the vast majority of 
which are believed to be AGN. 
This result, however, is strictly valid only at soft X-ray energies 
(1-3 keV) where imaging telescopes, necessary 
to carry out sensitive surveys, until very recently could operate.
The energy density of the XRB peaks at about 30 keV, so that the power 
emitted at 10 keV is about three times that emitted at 1 keV. 
The extrapolation of the low energy results to energies where 
most of the XRB power is emitted would in principle be straightforward 
if the spectra of the sources detected in the soft X-rays were the same 
as that of the XRB (i.e. $\alpha \sim 0.4$). It is well known 
that this is far from being the case, with a "canonical" 2-10 keV energy 
spectral index of bright AGN $\alpha~ \sim  0.7 $
and a much steeper average AGN soft X-ray spectral slope ($\alpha~ \sim 1.5$, 
Yuan et al. 1998), that probably hardens to $\alpha~ \sim 1.0 $ at 
very faint fluxes (Hasinger et al. 1993, Almaini et al. 1996). 
This frustrating situation, where AGN are detected in sufficient number 
to easily explain the XRB but with a wrong (or very wrong) spectral slope,
is usually referred to as the {\it spectral paradox}. 
Moreover, the unified schemes of AGN predict that many obscured AGN 
that cannot be easily (or at all!) detected in the ROSAT band could play 
an important role in the contribution to the XRB at higher energies 
(Setti \& Woltjer 1989, Madau et al. 1994, Comastri et al. 1995).
In a comparison between the soft X-ray logN-logS and the counts in the
\newpage \ni
2-10 keV band, it appeared that the extrapolation of the soft sources 
could only contribute about one third of the observed counts if the assumed 
spectral slope is $\alpha =1.0$. (e.g. Georgantopoulos et al. 1997, Cagnoni 
et al. 1998).
This has been interpreted as possible evidence for a new population 
of cosmic sources that could dominate the hard X-ray counts and therefore 
also make up most of the XRB.

\begin{figure}
\centerline{\psfig{file=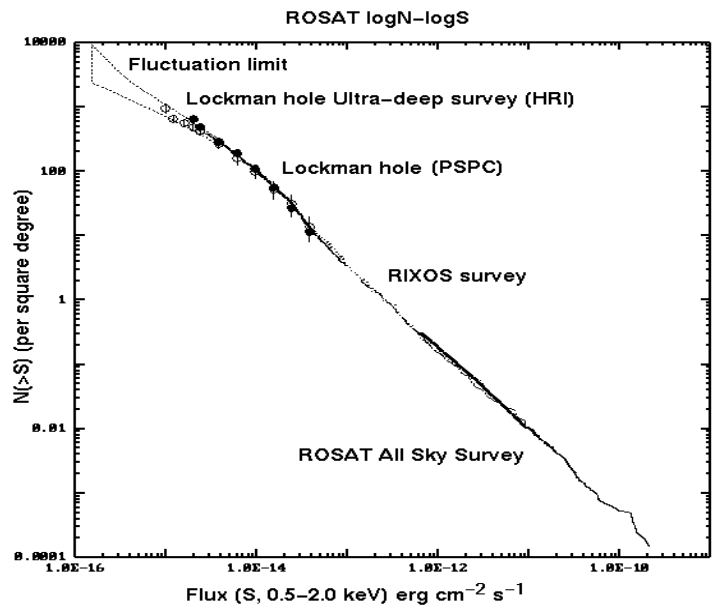, width=9cm, height=11cm, angle=-90}}
\caption{FIGURE 1. The total 0.5-2 keV logN-logS determined from several 
ROSAT surveys as indicated in the plot. 
Adapted from Hasinger et al. 1998}
\end{figure}

\ni     
\sk
\ni 2. SOFT X-RAY SURVEYS 
\ssk
\ni 
Recent findings from the deepest soft X-ray surveys have been reported in 
several papers (e.g. Hasinger et al. 1998, McHardy et al. 1998, 
Georgantopoulos et al. 1996, Bower et al. 1996, Boyle et al. 1995,
Branduardi-Raymont et al. 1994). Here we only give a brief summary of 
the results and refer the reader to the literature for details. 
The present knowledge in this field is condensed in 
Figure 1 which shows the total (that is including all sources regardless 
to their optical identification) 0.5-2.0 keV logN-logS assembled using 
several data sets. These include a) the ROSAT all sky survey bright source 
catalog (Voges et al. 1995), b) the RIXOS survey (Mason et al. 1998), c) 
the deep (207 ksec) PSPC pointing and d) the ultra-deep (1.1 Msec) 0.3 
$ deg ^2$ HRI survey of the Lockman Hole region (Hasinger et al. 1998). 
The integral number-count distribution is a steep function of flux 
until $ 2\times 10^{-14}$ \ergs where a 
break to a flatter than Euclidean slope is clearly present.
The counts, reaching a value of $ \sim 1000~deg^2$ at  
$S=1\times 10^{-15}$ \ergs imply that 70-80 \% of the XRB at 1 keV is 
resolved into discrete sources. Several programs for the identification 
of these X-ray sources are well under way. Results from the deepest surveys 
show that a large fraction of the faint sources are classical broad-line AGN, 
although the possibility that a significant fraction of narrow emission 
line galaxies (NELG) could be present (e.g. McHardy et al. 1998) is still 
debated. The identification program from the HRI ultra-deep survey, 
the deepest survey ever made, however, shows that the number of sources 
identified with NELG is small (Schmidt et al. 1998). 

\sk
\ni 3. ASCA AND BEPPOSAX SURVEYS IN HARDER X-RAY BANDS.
\ssk
\ni 
The many important achievements in the soft X-ray band contrast with 
the situation at slightly higher energies (2-10 keV) where until very recently 
only a tiny percentage ($\approx$ 3\%) of the XRB could be resolved 
into discrete sources.
This situation significantly improved when the first satellites 
carrying on board instruments capable of producing X-ray images at 
energies as large as 10 keV (ASCA and \sax) became operational. 
Several surveys have been carried out in the 2-10 keV band with ASCA. 
The most recent results are reported in Ueda et al. 1998 (Large Sky 
Survey, LSS) Ogasaka et al. 1998 (Deep Sky Survey, DSS), and 
Cagnoni et al. 1998, Della Ceca et al. 1998 (ASCA GIS Medium Survey),
Georgantopoulos et al. 1997, (ASCA deep observations of ROSAT deep fields).
The contributions of these surveys to the 2-10 keV logN-logS 
are shown in figure 2 together with the preliminary results from
the \sax deep surveys (see below).
About 30\% of the XRB is now resolved into discrete sources at 
$S \sim 5\times 10^{-14}$ \ergs. 
Optical identification programs have just started; however,
due to the relatively large ASCA position uncertainties for serendipitous 
sources, this process may require a long time. Initial results 
clearly indicate (Fiore et al., this volume) that samples selected in 
the 2-10 keV band include a quantitatively different mix of optical 
counterparts than that seen in the soft X-rays (e.g. Ueda et al. 1998). 

\sk
\ni 3.1 THE BEPPOSAX DEEP AND SERENDIPITOUS SURVEYS 
\ssk
\ni
The \sax deep surveys are an on-going project to analyze in a homogeneous way
the deepest X-ray images taken with the MECS instrument. At this moment the
analysis covers 16 high galactic latitude fields for a total exposure of 
1.3 million seconds. Initial results (Giommi et al 1998) are limited 
to the analysis of the central 8.5 arcminutes where the sensitivity is
relatively constant and no complications due to the window support structure 
are present. The area covered is therefore about one degree, with a 
sensitivity limit of $\approx 4\times 10^{-14}$ \ergs in the 2-10 keV band.
The resulting LogN-LogS is shown in figure 2 together with several other 
results obtained with ASCA, Ginga and HEAO1.
This project will be extended to the full MECS field of view and 
will also include serendipitous sources found in all public \sax fields. 
This type of approach has already started with the HELLAS survey in the 
hard part (5-10 keV) of the MECS sensitivity range (Fiore et al. 
this volume).   
So far the HELLAS survey includes data from 120 high Galactic latitude 
MECS fields.
Sources are searched in images accumulated between $\sim$ 4.5 and 10 keV 
using an improved version of the DETECT routine of the XIMAGE package 
(see Ricci et al. 1998 and Giommi et al. 1998 for details).  
Correction for the energy dependent vignetting and PSF are applied to the 
counts that are then extracted in three bands 4.5-10, 2.5-4.5 and 1.3-2.5 keV.
So far 120 MECS fields have been analyzed and 177 sources have been detected 
at a confidence level $> 3.5 \sigma $. The source fluxes range from 
$4 \times 10^{-14}$ to a few $ \times 10^{-12}$ \ergs in the 5-10 keV band.
The resulting cumulative LogN-LogS is shown in figure 2. 
\begin{figure}
\centerline{\psfig{file=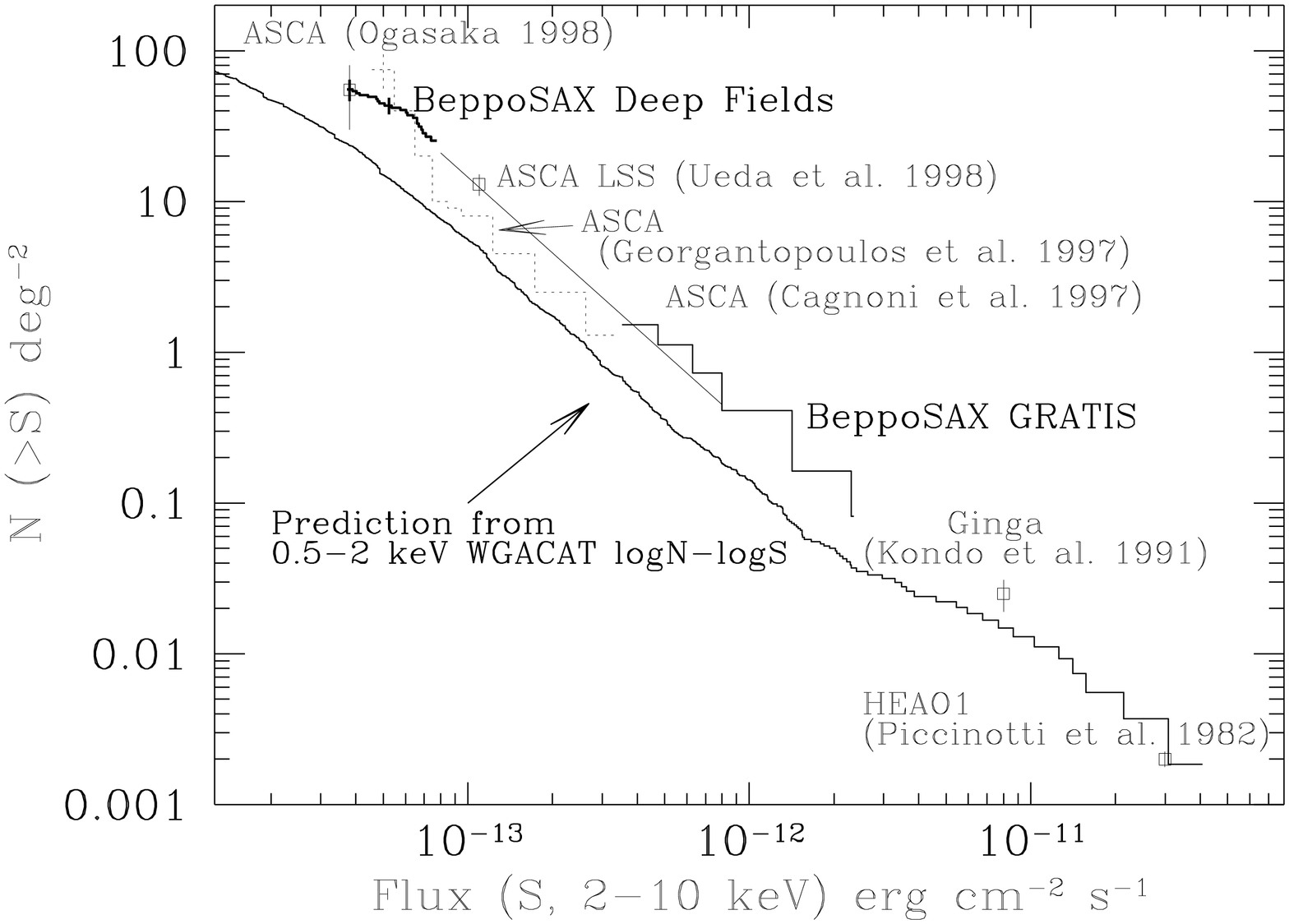, height=6.9cm,  angle=-90} 
\psfig{file=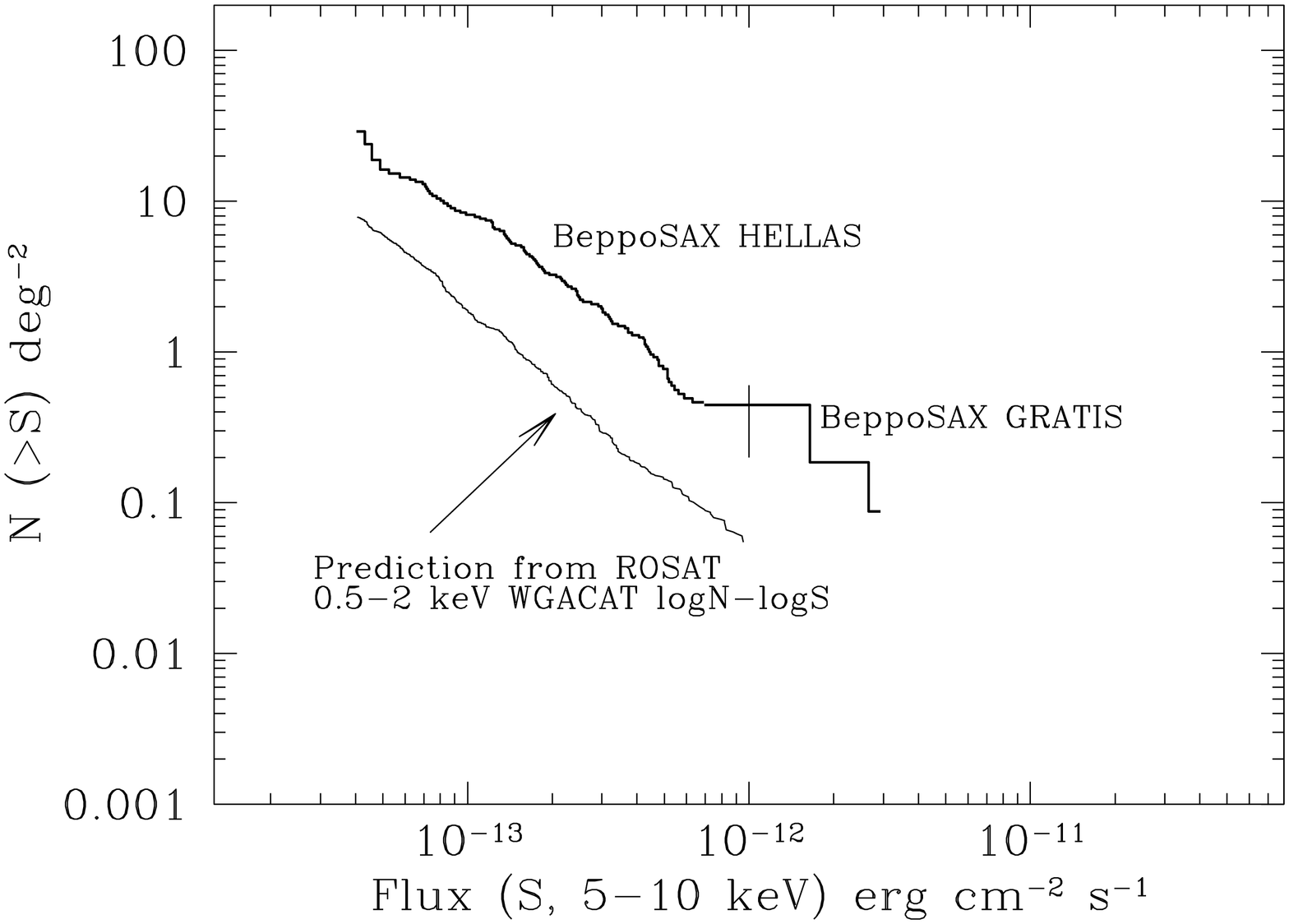, height=6.9cm, angle=-90}}
\sk
\caption{FIGURE 2. The 2-10 (left) and 5-10 keV (right) logN-logS functions
as determined from various \sax and ASCA surveys. The expected logN-logS 
derived from a sample of ROSAT WGA sources, taking into account the observed
spectral slopes, are also shown. In the 2-10 keV band the counts at bright 
fluxes (HEAO1-A2, Ginga points) are reasonably well predicted, in all other 
cases a clear deficit of a factor 2-3 is apparent.}
\end{figure}
\sk
\ni 3.2 THE BEPPOSAX GRATIS SURVEY
\ssk
\ni
The GRATIS Survey (GRand Area Target acquisition Intermediate 
pointings Survey, Perri et al. 1998) consists of all MECS short 
observations that are performed as part of the (single gyro and gyro-less) 
procedure to point the spacecraft from one target to the following.
Typical exposures are 4,000 seconds corresponding to a sensitivity of
$\sim $ 3-4 x $10^{-13}$ \ergs.
The logN-logS estimated from the 8 sources detected in an initial set 
of 32 distinct pointings covering a total of about 14 degrees of sky are shown 
in figures 2. At the end of the mission the GRATIS survey will cover about 
50 square degrees of high galactic latitude sky. Although its sensitivity 
limit is only moderate the GRATIS survey has the advantage of not suffering 
from incompleteness at fluxes comparable (or higher) than those of the 
targets; a bias that is hardly avoidable in serendipitous surveys.  
This survey is therefore a nice complement to the deep and 
serendipitous surveys and provides a fair estimation of the bright 
end of the logN-logS.

\sk
\ni 4. A COMPARISON BETWEEN SOFT AND HARD X-RAY SURVEYS 
\ssk
Extrapolations of the soft X-ray source counts to the harder 2-10 keV band 
have been done under the assumption that all sources are characterized by a 
single spectral slope, generally assumed to be  $\alpha =1.$ (e.g. 
Georgantopoulos et al.1997, Cagnoni et al. 1998, Ueda et al. 1998). 
While this a fair approximation 
at faint fluxes (Hasinger et al. 1993) it is certainly not a good one above 
$2-3\times 10^{-14} $ \ergs where the average source spectral slope is much 
steeper and the dispersion around the mean is large (Yuan et al. 1998).
This large dispersion is most probably present at all flux levels and may 
be the cause of large uncertainties. 
To minimize difficulties arising from these problems  
we have used the WGA catalog of ROSAT PSPC sources (White Giommi 
\& Angelini, 1994) to derive the LogN-LogS taking into account
the spectral slope of each detected source as estimated from its hardness ratio.
The sample includes about 9,000 sources extracted from the WGA catalog 
according to the following criteria: 
\ssk
- Distance from target $> 2.5$ arcminutes \sssk
- Off-Axis angle $< 11$ or  $25 < $ off-axis $ < 45$ 
arcminutes (to avoid the target area and problems with reduced
sensitivity due to window support structure + wobble) \sssk
- $N_H < 3\times 10^{20} cm^{-2} $\sssk
- exposure between 4,000 and 100,000 seconds \sssk
- exclude fields centered on clusters of galaxies (800000 $<$ ROR $<$ 899999) \sssk
- $-60 < Dec < +90$ (to avoid the Magellanic Clouds region) \sssk
\ssk
The detailed sky coverage for different assumptions of the power law energy 
index ($\alpha$) is shown in figure 3. Note the very large dependence
on $\alpha $, especially for steep spectra. 
To compare the resulting 0.5-2 keV logN-logS with previous results we have 
first assumed a single spectral slope $\alpha =1.$ as in Hasinger  
et al. 1998. Our logN-logS is in very good agreement with the counts shown 
in figure 1 up to $\sim 1\times 10^{-14} $ \ergs; this proves that the 
WGA catalog, which was not originally designed for this purpose, can be 
successfully used for statistical studies. 
At very faint fluxes problems connected 
with source confusion and incompleteness of the WGA catalog cause
an underestimation of about a factor 2 at $S(0.5-2 keV) = 5\times 10^{-15} $. 
We conclude that our logN-logS is a good representation of reality at 
fluxes larger than $1\times 10^{-14} $ \ergs.
Next we have estimated the 0.5-2 keV logN-logS calculating the X-ray flux 
taking into account the measured spectral slope of each source (when 
available) and assuming an absorption equal to the amount of Galactic $N_H$ 
along the line of sight. 
For sources near the detection limit, where no reliable estimation of 
the hardness ratio is available, we have taken Monte Carlo simulated spectral 
slopes drawn from a distribution equal to that observed in good 
signal-to-noise ratio detections. At fluxes below $ 2\times 10^{-14} $ 
\ergs we have assumed a gaussian distribution of spectral indices with a 
mean value of 1.0 and a dispersion of 0.4.  
The resulting 0.5-2 keV logN-logS is similar in shape to that of figure 1. 
but with a normalization that is about 25\% lower. 
We have then extrapolated our WGA logN-logS to the 2-10 keV and 5-10 keV bands 
again using the measured spectral slope of each source 
as described above. We have compared the resulting counts with data 
from various satellites, including ASCA and \sax. 
In the 2-10 keV band the counts at bright fluxes (HEAO1-A2, Ginga points) are 
reasonably well predicted, in all other cases a deficit of a factor 2-3 is 
apparent, especially in the 5-10 keV band (figure 2).
As a last exercise we have extrapolated the 0.5-2 keV logN-logS assuming 
that the spectrum of each source hardens above 2 keV
by a fixed value $ \Delta \alpha $. We find that for $\Delta \alpha = 0.5 $
both the 2-10 and the 5-10 keV counts are reasonably well matched.
\begin{figure}
\centerline{ \psfig{file=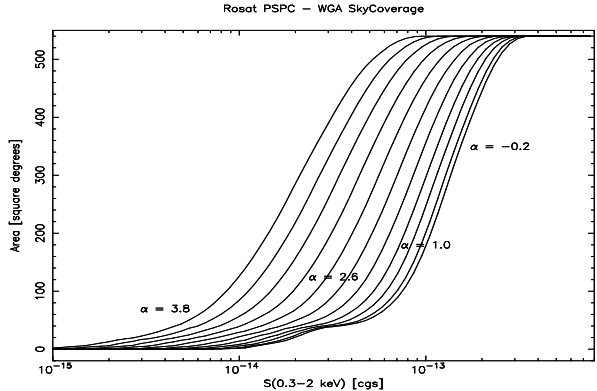, height=9cm, width=5.5cm, angle=-90} }
\caption{FIGURE 3. 
The sky coverage of the WGA sample for various assumptions of 
the assumed power law spectral slope. Note that the area covered at a given 
flux can be very different for different slopes.} 
\end{figure}
\begin{figure}
\centerline{\psfig{file=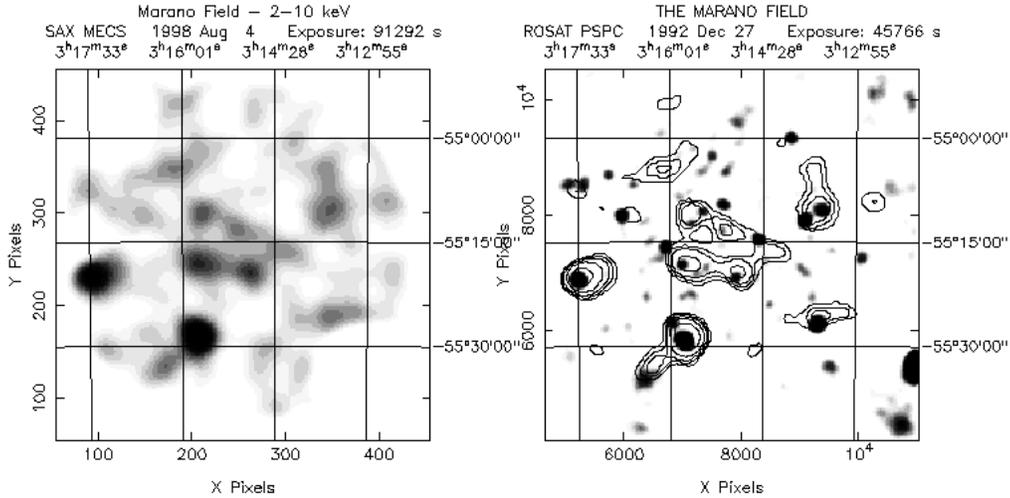, height=14.cm, angle=-90}}
\caption{FIGURE 4. MECS 2-10 keV background subtracted X-ray 
image of the Marano field. On the right side the 2-10 keV intensity contours are overlaid to a PSPC 0.1-2.4 keV image of the same field. 
Nearly all the MECS sources have a counterpart in the soft X-rays.}
\end{figure}
\ssk
\ni 5. BEPPOSAX LONG OBSERVATIONS OF ROSAT DEEP FIELDS
\ssk 
To perform a direct comparison of the sources detected in the soft X-ray band 
and at harder energies, some long ($> 100,000$ seconds) \sax observations 
have been done on fields that were previously studied in detail with ROSAT. 
These include a 130,000 seconds MECS exposure of the Lockman 
Hole (Tr\"umper et al. 1999), and a 91,000 seconds image of the Marano field 
(La Franca et al. 1999). In addition the \sax archive includes over 
100 long MECS exposures ($> 50,000$s) of PSPC fields. The full analysis 
of these data will be published elsewhere, here we only report some initial 
results. 
Figure 4 shows the \sax MECS background-subtracted image of 
the Marano field compared to a PSPC exposure of the same region of the sky 
taken from the public archive. 
From the MECS contours overlaid on the PSPC exposure we see that nearly 
all the sources detected in the 2-10 keV band are also detected in the 
PSPC image. A similar result has been obtained with the Lockman hole data. 
The systematic work on the comparison between \sax and ROSAT archival data 
is just started; however, we can preliminarily conclude that only a very 
small percentage of MECS sources is not detected in ROSAT deep fields. Part, 
or even all, of this small percentage of non-detections can be due to 
variability. These results imply that no large completely new population 
of sources is present above 2-3 keV. 
The predicted population of new or heavily cutoff sources, that was 
invoked by the early comparison between the soft and hard logN-logS 
and by XRB synthesis models, does not seem to show up in a straight-forward 
manner. 
One possible explanation for this is that the expected hard sources not 
only emit through the heavily cutoff nuclear component (hardly or not at 
all detectable by ROSAT) but also emit a small fraction of soft X-rays 
from non-nuclear components such a possible starburst region, or through 
reflection or partial transmission of nuclear X-rays. 
\ssk
\ni 6. CONCLUSIONS 
\ssk
\ni 
X-ray surveys are at a point where very important results have been obtained 
in the technologically simple, but astrophysically complex, soft X-ray band. 
Only recently significant achievements could be made at higher energies where 
most of the XRB power is emitted and where our view into the central parts 
of AGN becomes clearer.  
Early comparisons based on simple extrapolations of the soft X-ray counts
predicted the existence of a large population of previously unknown sources.
We have shown that the present data indicate that the situation is 
more complex. Taking into proper account the observed distribution of 
spectral slopes we see that the hard X-ray counts are not under-predicted 
if a moderate spectral hardening or a non-power law spectral shape is 
present. Direct comparisons between X-ray images show that most of the 
hard sources, many of which are expected to be heavily cutoff objects 
that should not be visible below 2-3 keV, are instead detected at 
moderately low soft X-ray fluxes.  
A possible explanation for these soft X-ray detections
is that a second component arising from a region external to the 
circumnuclear absorbing material (e.g. starburst, reflection component etc.) 
could make these sources detectable at soft energies. 
Another possible cause for the detection at soft energies could be
that a fraction of the nuclear radiation could escape from the
AGN nucleus unabsorbed because of partial covering of the circumnuclear 
material. 
These unforeseen sources of soft X-rays could affect 
the estimation of the luminosity function and the cosmological evolution 
of AGN. More data is however clearly necessary, and this will be abundantly 
available with the advent of the next generation of X-ray astronomy 
satellites.
}
\ssk
\baselineskip = 12pt
{\abstract \ni ACKNOWLEDGMENTS

We thank F. La Franca for allowing us to show the Marano Field \sax data in 
advance of publication.   
G. Hasinger and R. Della Ceca are thanked for communicating us some of their 
results prior to publication. 
We also would like to thank F. Pompilio for his contribution to the 
calculation of the \sax sky coverage.}

\baselineskip = 12pt


{\references \ni REFERENCES
\sssk
\ref Almaini O., et al. 1996, MNRAS, 282, 295 
\ref Bower R.G. et al. 1996, MNRAS 281, 59 
\ref Boyle B.J., McMahon G., Wilkes B.J., Elvis M., 1995 MNRAS 272, 462
\ref Branduardi-Raymont G., 1994, MNRAS, 272, 462
\ref Cagnoni, I., Della Ceca, R., and Maccacaro, T. 1998, APJ, 493, 54 
\ref Comastri, A., Setti, G., Zamorani, G., and Hasinger, G., 1995
A\&A, 296, 1
\ref Della Ceca, R., Cagnoni, I., and Maccacaro, T. 1998, 
Astronomische Nachrichten, 319, 64.
\ref Hasinger, G., et al. 1993 A\&A 275, 1
\ref Hasinger, G., 1998, Nuclear Physics B 4 (Proc. Suppl.) 69/1-3, 600
\ref Hasinger, G., et al. 1998, A\&A, 329, 482
\ref Georgantopoulos I., et al. 1996 MNRAS, 280, 276 
\ref Georgantopoulos I., et al. 1997 MNRAS, 291, 203 
\ref Giommi P. et al. 1998 Nuclear Physics B (proc. Suppl.) 69/1-3, 591
\ref La Franca, F. et al. 1999, in preparation
\ref Madau, P., Ghisellini, G., and Fabian, A., 1994, MNRAS, 270, L17
\ref Mason, K., et al. 1998, MNRAS, submitted 
\ref Mc Hardy et al. 1998, MNRAS, 295, 641 
\ref Ogasaka, Y., et al. 1998, Astronomische Nachrichten, 319, 43.
\ref Perri M., Giommi P., Fiore F., Antonelli L.A., 1998 Mem.S.A.It., in press
\ref Ricci D., et al. 1998 Nuclear Physics B 4 (Proc. Suppl.) 69/1-3, 618 
\ref Setti G., Woltjer L., 1989, A\&A 224, L21
\ref Schmidt, M., et al. 1998, A\&A, 329, 495
\ref Tr\"umper, J., Hasinger, G., Giommi, P. et al., 1999, in preparation
\ref Ueda, Y., et al. 1998, Nature, 391, 866.
\ref Voges, W., et al. 1995, IAU Circ \# 6102
\ref White, N. E., Giommi P. \& Angelini L. 1994. IAUC 6100 
\ref Yuan, W., Brinkmann, W., Siebert, J., and W. Voges, 1998, A\&A 330, 108 
}                      
\end{document}